\documentclass[aip,apl,reprint]{revtex4-1}
\usepackage[T1]{fontenc}
\usepackage[latin9]{inputenc}
\usepackage{amstext}
\usepackage{graphicx}
\usepackage{amssymb}
\usepackage{multirow}


\begin{document}

\preprint{}

\title{Surface loss simulations of superconducting coplanar waveguide resonators}

\author{J. Wenner}
\affiliation{Department of Physics, University of California, Santa Barbara, California 93106, USA}
\author{R. Barends}
\affiliation{Department of Physics, University of California, Santa Barbara, California 93106, USA}
\author{R. C. Bialczak}
\affiliation{Department of Physics, University of California, Santa Barbara, California 93106, USA}
\author{Yu Chen}
\affiliation{Department of Physics, University of California, Santa Barbara, California 93106, USA}
\author{J. Kelly}
\affiliation{Department of Physics, University of California, Santa Barbara, California 93106, USA}
\author{Erik Lucero}
\affiliation{Department of Physics, University of California, Santa Barbara, California 93106, USA}
\author{Matteo Mariantoni}
\affiliation{Department of Physics, University of California, Santa Barbara, California 93106, USA}
\author{A. Megrant}
\affiliation{Department of Physics, University of California, Santa Barbara, California 93106, USA}
\author{P. J. J. O'Malley}
\affiliation{Department of Physics, University of California, Santa Barbara, California 93106, USA}
\author{D. Sank}
\affiliation{Department of Physics, University of California, Santa Barbara, California 93106, USA}
\author{A. Vainsencher}
\affiliation{Department of Physics, University of California, Santa Barbara, California 93106, USA}
\author{H. Wang}
\affiliation{Department of Physics, University of California, Santa Barbara, California 93106, USA}
\affiliation{Department of Physics, Zhejiang University, Hangzhou 310027, China}
\author{T. C. White}
\affiliation{Department of Physics, University of California, Santa Barbara, California 93106, USA}
\author{Y. Yin}
\affiliation{Department of Physics, University of California, Santa Barbara, California 93106, USA}
\author{J. Zhao}
\affiliation{Department of Physics, University of California, Santa Barbara, California 93106, USA}
\author{A. N. Cleland}
\affiliation{Department of Physics, University of California, Santa Barbara, California 93106, USA}
\author{John M. Martinis}
\email{martinis@physics.ucsb.edu}
\affiliation{Department of Physics, University of California, Santa Barbara, California 93106, USA}
\date{\today}

\begin{abstract}
Losses in superconducting planar resonators are presently assumed to predominantly arise from surface-oxide dissipation, due to experimental losses varying with choice of materials. We model and simulate the magnitude of the loss from interface surfaces in the resonator, and investigate the dependence on power, resonator geometry, and dimensions. Surprisingly, the dominant surface loss is found to arise from the metal-substrate and substrate-air interfaces. This result will be useful in guiding device optimization, even with conventional materials.
\end{abstract}
\maketitle

Superconducting coplanar waveguide (CPW) resonators are critical elements in photon detection \cite{Schlaerth2008}, quantum computation \cite{DiCarlo2009,Altomare2010,Steffen2010,Ansmann2009}, and creating and decohering quantum photon states \cite{Wang2010,Wang2011}. Such applications are limited by the energy decay time. One prominent source of decoherence at low powers in resonators has previously been found to be two-level states (TLSs) on the various surfaces \cite{Gao2008,Wang2009,Barends2010,Sage2011,Wisbey2010}. Knowing the TLS locations is important for improving the resonators. Previous measurements and simulations suggested that the exposed metal surface (metal-air interface) is a crucial decoherence source \cite{Wang2009,Barends2010,Sage2011}, which has driven research in using non-oxidizing superconductors for quantum devices. However, we show with simulations and a model that, for typical metal oxide parameters, the more likely source of CPW loss is instead the metal-substrate and substrate-air interfaces, thereby changing the approach needed to reduce losses.

\begin{figure}
\begin{center}
\includegraphics[width=3.25in]{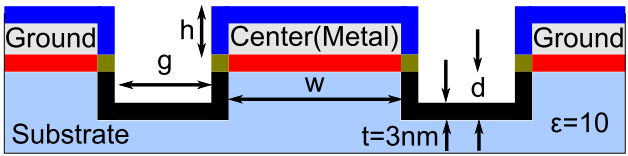}
\end{center}
\caption{(Color online) Coplanar waveguide dimensions and interfaces. Dimensions are the center width $w$, gap width $g$, metal height $h$, etch depth $d$, and assumed interface thickness $t$=3\,nm. This structure has metal-air (blue), metal-substrate (red), and substrate-air (black) interfaces. The 3\,nm$\times$3\,nm corner square (yellow) is treated separately.}
\label{FigCoplanarDims}
\end{figure}

Thin-film resonators have three types of amorphous interfaces that can contain TLSs and thus introduce surface losses (Fig.\,\ref{FigCoplanarDims}): metal-air (ma), metal-substrate (ms), and substrate-air (sa). Each interface can include a thin oxide or contaminant layer sandwiched between the two primary layers. The total loss tangent for these thin layers is $\sum_i p_i\tan\delta_i$, where surface interface type $i$ has loss tangent $\tan\delta_i$ and participation ratio \cite{Koch2007,Gao2008}
\begin{equation}
p_i=W^{-1}t_i\epsilon_i\int ds\,|E|^2,
\label{EqParticipationDef}
\end{equation}
for thickness $t_i$, dielectric constant $\epsilon_i$, length coordinate $s$, and energy per unit length $W$. Using the boundary conditions on the electric displacement gives the three interface participation ratios \cite{Supp}:
\begin{eqnarray}
p_\text{ma}W/t_\text{ma} &=& \epsilon_\text{ma}^{-1}\int ds\,|E_{\text{a}\perp}|^2 \label{EqRatioMA}\\
p_\text{ms}W/t_\text{ms} &=& (\epsilon_\text{s}^2/\epsilon_\text{ms})\int ds\,|E_{\text{s}\perp}|^2 \label{EqRatioMS}\\
p_\text{sa}W/t_\text{sa} &=& \epsilon_\text{sa}\int ds\,|E_{\text{a}\parallel}|^2 +\epsilon_\text{sa}^{-1}\int ds\,|E_{\text{a}\perp}|^2 \label{EqRatioSA},
\end{eqnarray}
where $E_\text{a}$ ($E_\text{s}$) is the electric field in the air (substrate) outside the interface and $E_\parallel$ ($E_\perp$) is the electric field component parallel (perpendicular) to the interface.

Here, we take all dielectric constants to be of order $\epsilon\sim10$, typical of metal oxides. Then, $p_\text{ma}$ and $p_{\text{sa},\perp}$ are of order 1\% of $p_\text{ms}$ and $p_{\text{sa},\parallel}$. Thus, if all interface loss tangents and thicknesses are similar, the substrate-air and metal-substrate interfaces are 100 times more lossy than the metal-air interface.

\begin{table}
\caption{Simulated losses for three resonator geometries \cite{Supp}. Losses are calculated for $\epsilon_s=10$ and surface dielectrics with $\epsilon = 10$, $t$=3\,nm, and loss tangent 0.002; dimensions are as in Fig.\,\ref{FigCoplanarDims}. Both coplanar waveguide (CPW) geometries, with identical dimensions, have similar metal-air (ma), metal-substrate (ms), and substrate-air (sa) losses, but etching the exposed substrate (Etched CPW) substantially reduces corner (c) loss. A microstrip geometry (with dielectric height $s$) has significantly less ma and sa losses than CPWs.}
\begin{tabular}{|l|c|c|c|c||c|c|c|c|}
\hline
\hline
 & \multicolumn{4}{|c||}{Dimensions ($\mu$m)} & \multicolumn{4}{|c|}{Loss $\times 10^6$} \\
\cline{2-9}
Type & $w$ & $h$ & $g$ & $d$ & ma & ms & sa & c \\
\hline
\hline
CPW & 5 & 0.1 & 2 & 0 & 0.10 & 6.13 & 4.02 & 1.32\\
\hline
Etched CPW & 5 & 0.1 & 2 & 0.01 & 0.11 & 4.64 & 5.25 & 0.39 \\
\hline
Microstrip & 20 & 0.2 & $s$=2 & 0 & 0.02 & 6.60 & 0.79 & 0.38 \\
\hline
\hline
\end{tabular}
\label{TabGeometries}
\end{table}

To accurately compare to this model, the participation ratios were numerically calculated using the finite-element solver COMSOL\cite{COMSOL}, with Eqs.\,(\ref{EqRatioMA})-(\ref{EqRatioSA}) used to extract the participation ratios from surface fields \cite{Supp}. As shown in Table \ref{TabGeometries}, $p_\text{ma}$ is 40-60 times smaller than $p_\text{ms}$ and $p_\text{sa}$, which corresponds to the difference in prefactors in Eqs.\,(\ref{EqRatioMA})-(\ref{EqRatioSA}), validating the discussion above. In addition, for typical interface parameters ($t\sim3$\,nm\cite{Wang2009}, $\tan\delta\sim0.002$\cite{OConnell2008}), the metal-air interface gives a quality factor $Q\sim10^7$, much greater than typically measured values of $10^5$; the measured values are more similar to the values of $1.5-2.5\times10^5$ predicted for the metal-substrate and substrate-air interfaces.

\begin{figure}[b]
\begin{center}
\includegraphics[width=3.25in]{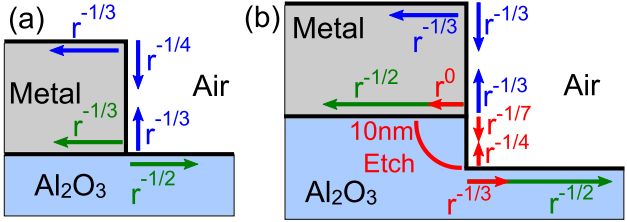}
\end{center}
\caption{(Color online) Electric field scaling on distances for CPW resonators. (\textbf{a}) is for an ordinary CPW geometry, while (\textbf{b}) is for a CPW with the substrate etched in the coplanar gap. Arrows indicate for which distance $r$ from the corner the electric field has the indicated scaling. As expected \cite{Jackson}, the field far from metal edges (b) scales as $r^{-1/2}$ and fields at the metal edge in both cases scale as $r^{-1/3}$; a substantial fraction of the energy in both cases is concentrated at the corners. Etching the substrate in the coplanar gap reduces this dependence over a length corresponding to the etch depth, reducing the corner participation ratio.}
\label{FigFieldDependence}
\end{figure}

Participation ratios have dominate contributions from the edges around the coplanar gap\cite{Supp}; the scaling of the electric fields with distance is displayed in Fig.\,\ref{FigFieldDependence}. For distances $r$ from corners much greater than other relevant dimensions, the field scales as $r^{-1/2}$, the predicted scaling for the field from a flat edge \cite{Jackson}. The electric fields on the metal-air interface and on the unetched metal-substrate interface scale as $r^{-1/3}$, the predicted behavior for the field from a metal 90$^\circ$ corner \cite{Jackson}; in this case, 20\% of the participation ratio is within 1\% of the length at the edge.  Because of this, we also calculate the participation ratio $p_\text{c}$ for a 3\,nm square at metal-air-substrate corners (as shown in Fig.\,\ref{FigCoplanarDims}), and find the loss from this small corner is only 3-4 times smaller than from the metal-air and substrate-air interfaces.

Etching the substrate in the coplanar gap flattens the field dependence on $r$ at the metal-air-substrate corner (Fig.\,\ref{FigFieldDependence}). This implies etching the substrate significantly reduces $p_\text{c}$ while leaving $p_\text{ma}$ and $p_\text{ms}$ unchanged, along with a potential decrease in $p_\text{sa}$ due to lower surface fields. Remarkably, even a 10\,nm etch reduces $p_\text{c}$ by 70\%, while a 2\,$\mu$m etch reduces $p_\text{c}$ by 99\% along with $p_\text{sa}$ by 50\%.

The microstrip geometry also changes the participation ratios, as shown in Table \ref{TabGeometries}.  Compared to CPW resonators, this approach significantly reduces $p_\text{c}$, $p_\text{ma}$, and $p_\text{sa}$ while leaving $p_\text{ms}$ unchanged (Table \ref{TabGeometries}); hence, microstrip resonators are especially useful if the metal-substrate interface is the least lossy interface. This difference between the interfaces implies that it is possible to determine if the metal-substrate interface is dominant by comparing losses from CPW and microstrip resonators.

\begin{figure}
\begin{center}
\includegraphics[width=3.25in]{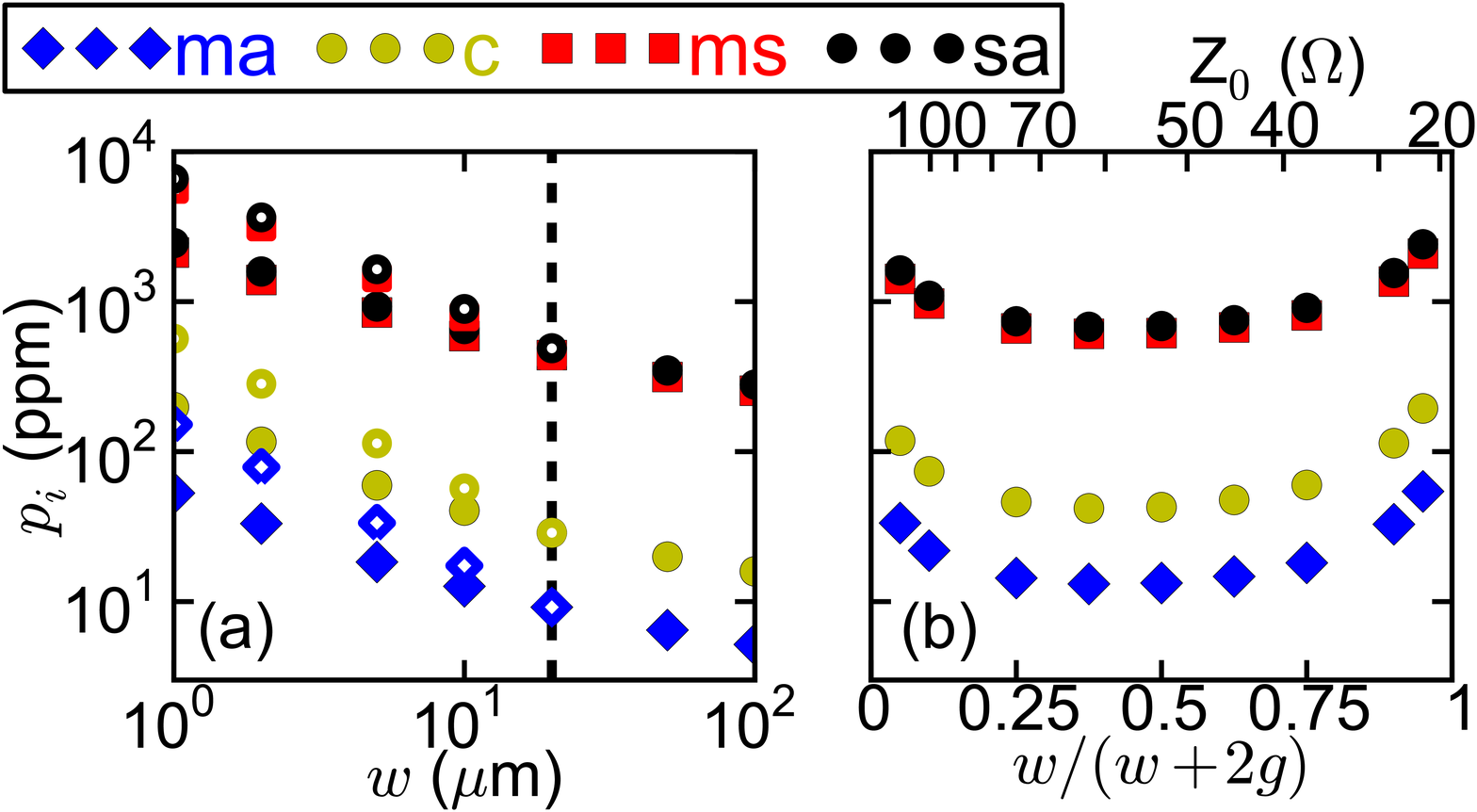}
\end{center}
\caption{(Color online) Geometric dependence of participation ratios for the metal-air (ma), corner (c), metal-substrate (ms), and substrate-air (sa) interfaces.  We assume 3\,nm surface dielectrics with $\epsilon_\text{ma} = \epsilon_\text{ms} = \epsilon_\text{sa}=10$ along with $h$=100\,nm and $d$=10\,nm. (\textbf{a}) Open symbols are for $g=w$, where the loss decreases as $1/w$. Filled symbols are for fixed $g=$20\,$\mu$m (indicated by dashed line), where the loss decreases as $w^{-2/3}$. (\textbf{b}) Plot of loss vs. $w$ for fixed $w+2g$=40\,$\mu$m. Minimum loss occurs at a characteristic impedance $Z_0=$50-60\,$\Omega$. In all three cases, the participation ratios scale together, implying that changing geometrical parameters can not determine which interface is dominant.}
\label{FigGeometry}
\end{figure}

One might expect varying CPW resonator dimensions, as shown in Fig.\,\ref{FigGeometry}, would determine dominant interfaces. For a fixed coplanar gap $g$, the loss decreases as $w^{-2/3}$ for $w\leq g$ and flattens off for $w>g$. Since larger widths increase loss by producing slotline modes, increasing radiation, and allowing trapped flux in the center strip \cite{Song2009}, we characterized the case $w=g$, where loss is proportional to $1/w$. If the distance $w+2g$ between the ground planes is kept fixed, then minimal loss occurs at $Z_0=50-60\,\Omega$. However, in all three cases, the participation ratios for all four interface types have nearly the same dependence. Hence, varying dimensions can reduce loss but makes determining the dominant interface difficult.

\begin{figure}
\begin{center}
\includegraphics[width=3.25in]{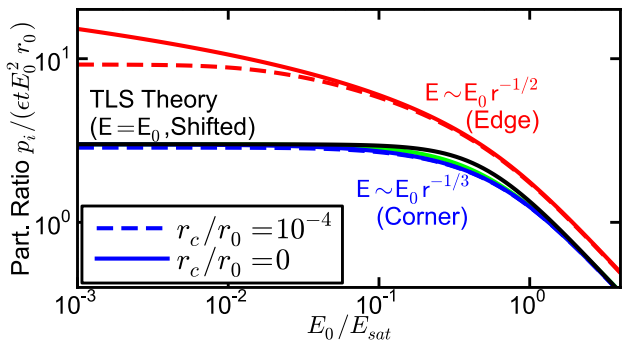}
\end{center}
\caption{(Color online) Plot of surface participation ratio vs. electric field for different field dependences. The case $E\sim r^{-1/3}$, for a metal corner [Eq.\,(\ref{EqCornerFull}), blue], has similar dependences if $r_c=0$ (solid) or if $r_c/r_0=10^{-4}$ (dashed); both are similar to the TLS saturation theory case [$E=E_0$, black] and the simplified formula [Eq.\,(\ref{EqApproxCorner}) for $\alpha=1.5$, green]. For the case $E\sim r^{-1/2}$, for a metal edge [Eq.\,(\ref{EqEdgeFull}], red), the $r_c/r_0=10^{-4}$ curve (dashed) has a broader crossover, while the $r_c=0$ (solid) case logarithmically diverges.}
\label{FigSaturation}
\end{figure}

One potential way to determine the key interfaces is by measuring the power dependence of the loss \cite{Barends2010}, since different interfaces have different field dependences from their corners (Fig.\,\ref{FigFieldDependence}). For a $90^\circ$ corner, where $E=E_0(r/r_0)^{-1/3}$, the surface participation ratio is \cite{Supp}
\begin{equation}
\frac{p}{t\epsilon}=3E_0^2r_0\left[\sqrt{1+\frac{E_0^2}{E_s^2}}-\sqrt{\left(\frac{r_c}{r_0}\right)^{2/3}+\frac{E_0^2}{E_s^2}}\right],
\label{EqCornerFull}
\end{equation}
where $E_s$ is the saturation field, $r_0$ is a characteristic length, and $r_c$ is a lower cutoff. For a metal edge, where $E=E_0(r/r_0)^{-1/2}$, the surface participation ratio is \cite{Supp}
\begin{equation}
\frac{p}{t\epsilon}=2E_0^2r_0\log\frac{1+\sqrt{1+\frac{E_0^2}{E_s^2}}}{\sqrt{\frac{r_c}{r_0}}+\sqrt{\frac{r_c}{r_0}+\frac{E_0^2}{E_s^2}}},
\label{EqEdgeFull}
\end{equation}
which gives a logarithmic divergence.  Comparing these results in Fig.\,\ref{FigSaturation}, we find the $r^{-1/2}$ edge model has a much broader crossover with the the drive field, while the $r^{-1/3}$ corner model is much closer to the sharp crossover of the simple TLS theory \cite{Wang2009}. These models are also well described (Fig.\,\ref{FigSaturation}) by the experimentally-based fitting formula \cite{Wang2009}
\begin{equation}
\frac{p}{t\epsilon}=\frac{cE_0^2r_0}{[1+0.9c(E_0/E_s)^\alpha]^{1/\alpha}},
\label{EqApproxCorner}
\end{equation}
where $c=3[1-(r_c/r_0)^{1/3}]$. Here, $\alpha\simeq1.5$ for the $r^{-1/3}$ case, similar to an experimentally-determined relation \cite{Wang2009}, while $\alpha\simeq0.75$ for the $r^{-1/2}$ case; this indicates that a corner, not an edge, dominated the experimental loss.

The surface loss model is partially validated by qualitatively explaining experimental results observed by other groups. For instance, HF-terminating a silicon substrate before growing an Nb film reduced the TLS loss by a factor of three compared to \textit{in situ} RF cleaning, confirming the importance of the metal-substrate interface \cite{Wisbey2010}. While the roughness in the coplanar gap did not affect their TLS loss, they substantially overetched into the substrate, so the substrate-air roughness experienced substantially lower fields. In addition, it was found that etching into the substrate reduced loss \cite{Barends2010}, confirming the importance of the substrate-air interface.

While previous experiments suggested that the metal-air interface is dominant, these results can be reexplained in terms of the model presented here. Our group previously found MBE-grown Re resonators had lower loss than sputter-grown Al resonators \cite{Wang2009}, which was attributed to the metal-air interface. This may instead be from differences in surface preparation and cleanliness between the growth conditions, affecting the metal-substrate interface. In addition, Nb resonators on Si and sapphire (Al$_2$O$_3$) substrates had similar losses and were more lossy than Al, Re, or TiN resonators; this was claimed to be from oxide formation at the metal-air interface \cite{Sage2011}. However, their Nb-Al$_2$O$_3$ resonators were grown on A-plane Al$_2$O$_3$, with which oxygen atoms diffuse into the Nb film \cite{Surgers1991}, making the metal-substrate interface significantly thicker and thus more lossy. In fact, others showed that Nb resonators on R-plane Al$_2$O$_3$ have losses similar to Re and Al \cite{Lindstrom2009}, indicating that the loss was not from oxide at the metal-air interface.

This model is not even restricted to CPW resonators. The Nb cavity resonators of Raimond, Haroche, and Brune \cite{Gleyzes2007} have a loss of $1/Q=2.4\times10^{-11}$. For typical parameters, the surface loss model gives a loss $\sim \tan\delta_\text{ma}p_\text{ma}/\epsilon_\text{ma}=2.2\times10^{-11}$, close to the experimental data.

In conclusion, we have developed a model for the resonator loss from interfaces. We find that, for realistic values, the metal-substrate and substrate-air interfaces are dominant, with participation ratios of order 100 times that of the metal-air interface. The loss can therefore be reduced by improving the metal-substrate and substrate-air interfaces, using microstrips with clean dielectrics, and increasing dimensions.

This work was supported by IARPA under ARO award W911NF-09-1-0375. M.M. acknowledges support from an Elings Postdoctoral Fellowship.

\end{document}


\preprint{}

\title{Supplementary Material for ``Surface loss simulations of superconducting coplanar waveguide resonators''}

\author{J. Wenner}
\affiliation{Department of Physics, University of California, Santa Barbara, CA 93106, USA}
\author{R. Barends}
\affiliation{Department of Physics, University of California, Santa Barbara, CA 93106, USA}
\author{R. C. Bialczak}
\affiliation{Department of Physics, University of California, Santa Barbara, CA 93106, USA}
\author{Yu Chen}
\affiliation{Department of Physics, University of California, Santa Barbara, CA 93106, USA}
\author{J. Kelly}
\affiliation{Department of Physics, University of California, Santa Barbara, CA 93106, USA}
\author{Erik Lucero}
\affiliation{Department of Physics, University of California, Santa Barbara, CA 93106, USA}
\author{Matteo Mariantoni}
\affiliation{Department of Physics, University of California, Santa Barbara, CA 93106, USA}
\author{A. Megrant}
\affiliation{Department of Physics, University of California, Santa Barbara, CA 93106, USA}
\author{P. J. J. O'Malley}
\affiliation{Department of Physics, University of California, Santa Barbara, CA 93106, USA}
\author{D. Sank}
\affiliation{Department of Physics, University of California, Santa Barbara, CA 93106, USA}
\author{A. Vainsencher}
\affiliation{Department of Physics, University of California, Santa Barbara, CA 93106, USA}
\author{H. Wang}
\affiliation{Department of Physics, University of California, Santa Barbara, CA 93106, USA}
\affiliation{Department of Physics, Zhejiang University, Hangzhou 310027, China}
\author{T. C. White}
\affiliation{Department of Physics, University of California, Santa Barbara, CA 93106, USA}
\author{Y. Yin}
\affiliation{Department of Physics, University of California, Santa Barbara, CA 93106, USA}
\author{J. Zhao}
\affiliation{Department of Physics, University of California, Santa Barbara, CA 93106, USA}
\author{A. N. Cleland}
\affiliation{Department of Physics, University of California, Santa Barbara, CA 93106, USA}
\author{John M. Martinis}
\email{martinis@physics.ucsb.edu}
\affiliation{Department of Physics, University of California, Santa Barbara, CA 93106, USA}

\date{\today}

\renewcommand{\thefigure}{S\arabic{figure}} 
\renewcommand\theequation {S\arabic{equation}}
\renewcommand{\thetable}{S\arabic{table}} 
\renewcommand\thepage {S\arabic{page}}
\setcounter{figure}{0}
\setcounter{equation}{0}
\setcounter{table}{0}
\setcounter{page}{1}

\begin{abstract}
Calculations are provided for the equations in the manuscript ``Surface loss simulations of superconducting coplanar waveguide resonators''. We provide a table of surface loss participation ratios for different geometries.
\end{abstract}
\maketitle

\section{Derivation of Surface Loss Model}

In Eq.\,(1) of the main manuscript, the participation ratio for interface $i$ is given by \cite{Koch2007,Gao2008}
\begin{equation}
p_i=W^{-1}t_i\epsilon_i\int ds\,|E|^2,
\label{SuppEqParticipationDef}
\end{equation}
where the interface has a small thickness $t_i$, dielectric constant $\epsilon_i$, and length coordinate $s$ and where the resonator structure has an energy per unit length $W$.

The metal-air (ma) interface consists of the metal, a thin metal oxide with thickness $t_\text{ma}\simeq3$\,nm and dielectric constant $\epsilon_\text{ma}$, and the outer air (vacuum) with $\epsilon_\text{a} = 1$. The electric field must be perpendicular to the metal surface, and because the interface layer is thin, we also approximate it as perpendicular in the dielectric, so $E_\text{ma}=E_{\text{ma}\perp}$. The continuity of $\epsilon E$ at the metal-oxide and air interface requires $\epsilon_\text{ma} E_{\text{ma}\perp,t} = E_{\text{a}\perp,t}$. Since the oxide is thin, $E$ does not change significantly over the oxide thickness. Combining all these results gives $E_\text{ma}\approx E_{\text{a}\perp}/\epsilon_\text{ma}$, so the participation ratio of the metal-air oxide is
\begin{eqnarray}
p_\text{ma}W/t_\text{ma} &=& \epsilon_\text{ma} \int ds\ |E_\text{ma}|^2 \nonumber \\
&=& \epsilon_\text{ma} \int ds\ |E_{\text{a}\perp}/\epsilon_\text{ma}|^2 \nonumber \\
&=& \epsilon_\text{ma}^{-1} \int ds\ |E_{\text{a}\perp}|^2.\label{SuppEqMA}
\end{eqnarray}

For the metal-substrate interface, we assume a thin dielectric layer of unknown origin between the metal and substrate, which might arise from a chemical reaction of the metal to the substrate or chemi- or physi-sorbed water on the wafer surface. As before, the electric field is perpendicular to the metal and the continuity of the displacement field requires $\epsilon_\text{ms} E_{\text{ms}\perp,t} = E_{\text{s}\perp,t}$, where ms represents this dielectric and s the substrate. Thus, we find $E_\text{ms}\approx E_{\text{s}\perp}\epsilon_\text{s}/\epsilon_\text{ms}$, so the participation ratio of the metal-substrate layer is
\begin{eqnarray}
p_\text{ms}W/t_\text{ms} &=& \epsilon_\text{ms} \int ds\ |E_\text{ms}|^2 \nonumber \\
&=& \epsilon_\text{ms} \int ds\ |E_{\text{s}\perp}\epsilon_\text{s}/\epsilon_\text{ms}|^2 \nonumber \\
&=& (\epsilon_\text{s}^2/\epsilon_\text{ms}) \int ds\ |E_{\text{s}\perp}|^2.\label{SuppEqMS}
\end{eqnarray}

For the substrate-air interface, there can be a dielectric layer from surface water or other contaminants from the air, described by a subscript sa. In addition to the perpendicular electric field which obeys $E_\text{sa}\approx E_{\text{a}\perp}/\epsilon_\text{sa}$ as before, there are also parallel field components obeying the boundary condition $E_{\text{a}\parallel}= E_{\text{sa}\parallel}=E_{\text{s}\parallel}$, since the interface layer is thin. Hence, the participation ratio of the substrate-air interface layer is
\begin{eqnarray}
p_\text{sa}W/t_\text{sa} &=& \epsilon_\text{sa} \int ds\ \left( |E_{\text{sa}\parallel}|^2 +|E_{\text{sa}\perp}|^2 \right) \nonumber \\
&=& \epsilon_\text{sa} \int ds\ |E_{\text{a}\parallel}|^2 + \epsilon_\text{sa}^{-1} \smallint ds\ |E_{\text{a}\perp}|^2.\label{SuppEqSA}
\end{eqnarray}

\section{Simulation Approach}

The coplanar and microstrip structures were simulated using the electric quasi-statics component of the finite element solver COMSOL's AC/DC module \cite{COMSOL}. We simulated a two dimensional cross-section with half of the resonator, using symmetry to account for the other half. We used adaptive meshing as a starting point and then performed additional meshing around the edges and the corners.

\begin{figure}[b]
\begin{center}
\includegraphics[width=3.25in]{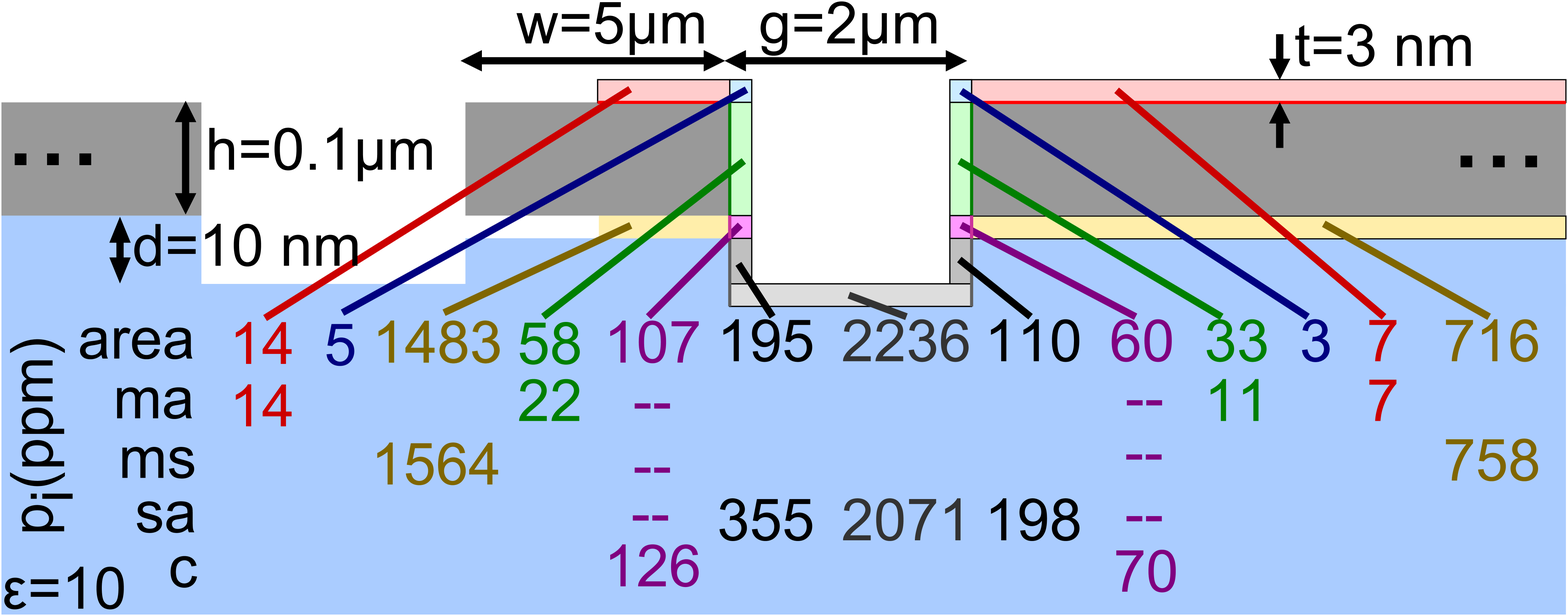}
\end{center}
\caption{Etched coplanar waveguide participation ratios. Participation ratios are given in parts per million (ppm) for the metal-air (ma), metal-substrate (ms), and substrate-air (sa) interfaces, as calculated with the surface fields approach with the metal-air-substrate corner (c) treated separately. Participation ratios for the area approach are shown separately. We assumed 3\,nm surface dielectrics with $\epsilon_\text{ma} = \epsilon_\text{ms} = \epsilon_\text{sa}=10$; the geometry is given in Table \ref{tab:loss} in cases c1 and c2. The participation ratios $p_\text{ma}+p_\text{c}$, $p_\text{ms}$, and $p_\text{sa}$ agree to within 15\%, validating the simulations against numerical errors.}
\label{FigAreaSurface}
\end{figure}

\begin{table*}[t]
\caption{\label{tab:loss} Simulation results for a variety of microwave resonators, obtained from the primary surface-based model [s], the area model [a], and a previous calculation \cite{Wang2009}. Coplanar waveguide and microstrip resonator dimensions are as indictated in Fig.\,\ref{FigDims}; the angle $\theta$ enclosed by the metal at the metal-substrate-air corner is assumed to be $90^\circ$ unless otherwise noted. The + sign in the metal-air column data is for a 3\,nm by 3\,nm area (c) at the intersection of the metal-air (ma), metal-substrate (ms), and substrate-air (sa) interfaces, and represents an entry that could be split among the three interface types. It is placed in the metal-air column since there it gives the greatest proportional uncertainty and because $p_\text{ma}+p_\text{c}$ and not $p_\text{ma}$ alone is comparable between the area and surface models. We assume $\epsilon_\text{s}=10$ and surface dielectrics with $\epsilon = 10$, thickness 3\,nm, and loss tangent 0.002.}
\begin{tabular}{llccccccc}
\hline
\hline
type \ \ \ \ \ \ \ \ \  & dimensions\  & capacitance &
\ \ metal-air \ & metal-sub. \ & \ sub.-air \  &
\ \ loss metal-air \ & \ loss metal-sub. \ & \ loss sub.-air \  \\
 & ($\mu$m) & pF/m & $p_\text{ma}$ (ppm) & $p_\text{ms}$ (ppm) & $p_\text{sa}$ (ppm) & $\times 10^6$ & $\times 10^6$ & $\times 10^6$ \\
\hline
\hline
coplanar & $w, h, g, d$ \\
\hline
c1 [a]  & 5, 0.1, 2, 0.01    & 162  & 119+167 & 2200  & 2541 & 0.24+0.33   & 4.40 & 5.08 \\
\hline
c2 [s]  & 5, 0.1, 2, 0.01    & 162  & 56+196  & 2322  & 2624 & 0.11+0.39   & 4.64 & 5.25 \\
\hline
c3 [a]  & 5, 0.1, 2, 0       & 163  & 290+387 & 2234  & 2286 & 0.58+0.77   & 4.47 & 4.57 \\
\hline
c4 [s]  & 5, 0.1, 2, 0       & 163  & 52+662  & 3065  & 2011 & 0.10+1.32   & 6.13 & 4.02 \\
\hline
c5 [a\cite{Wang2009}] & 5, 0.1, 2, 0 &  & 600 &       & 2000 & 1.2         &      & 4.0  \\
\hline
c6 [s]  & 5, 0.1, 2, 2       & 104  & 44+6    & 2690  & 1032 & 0.09+0.01   & 5.38 & 2.06 \\
\hline
c7 [s]  & 5, 0.025, 2, 0.01  & 161  & 55+209  & 2376  & 2735 & 0.11+0.42   & 4.75 & 5.47 \\
\hline
c8 [s]  & 2, 0.1, 20, 0.01   & 68   & 33+111  & 1394  & 1594 & 0.07+0.22   & 2.79 & 3.19 \\
\hline
c9 [s]  & 5, 0.1, 20, 0.01   & 85   & 18+60   & 847   & 928  & 0.04+0.12   & 1.69 & 1.85 \\
\hline
c10 [s] & 5, 0.1, 20, 0      & 85   & 17+207  & 1091  & 764  & 0.03+0.41   & 2.18 & 1.53 \\
\hline
c11 [s] & 5, 0.1, 20, 0,$\theta=45^\circ$      & 169   & 32+1414  & 3727  & 2267  & 0.06+2.83   & 7.45 & 4.53 \\
\hline
c12 [s] & 5, 0.1, 20, 0,$\theta=135^\circ$     & 158   & 104+695  & 2841  & 1963  & 0.21+1.39   & 5.68 & 3.93 \\
\hline
\hline
microstrip & $w, h, s, d$ \\
\hline
m1 [s]  & 20, 0.2, 2, 0.01   & 985  & 10+45   & 3155  & 526  & 0.02+0.09   & 6.31 & 1.05 \\
\hline
m2 [s]  & 20, 0.2, 0.2, 0.01 & 8964 & 7+55    & 29942 & 409  & 0.01+0.11   & 59.9 & 0.82 \\
\hline
m3 [s]  & 10, 0.2, 2, 0.01   & 539  & 19+82   & 3301  & 964  & 0.04+0.16   & 6.60 & 1.93 \\
\hline
m4 [s]  & 20, 0.02, 2, 0.01  & 983  & 10+49   & 3185  & 557  & 0.02+0.10   & 6.37 & 1.11 \\
\hline
m5 [s]  & 20, 0.2, 2, 0      & 987  & 9.3+189 & 3301  & 397  & 0.02+0.38   & 6.60 & 0.79 \\
\hline
m6 [s]  & 20, 0.2, 2, 2      & 914  & 4.6+1.9 & 2924  & 291  & 0.009+0.004 & 5.85 & 0.58 \\
\hline
m7 [s]  & 20, 0.2, 2, -2     & 1006 & 1.5+3.2 & 3192  & 241  & 0.003+0.006 & 6.38 & 0.48 \\
\hline
\hline
\end{tabular}
\end{table*}

To determine the participation ratios, we initially treated the interfaces as 3\,nm thick dielectrics with dielectric constant $\epsilon=10$. However, this area approach is computationally expensive since it requires meshing on the nanometer scale over distances of hundreds of microns. As such, we primarily calculated the participation ratios by computing the electric field on all boundary interfaces with the interface dielectrics excluded from the model and then applying Eqs.\,(\ref{SuppEqMA})-(\ref{SuppEqSA}). This is less computationally expensive as there is no thin dielectric layer explicitly included at the interfaces which needs to be carefully meshed. As indicated in Fig.\,\ref{FigAreaSurface} and Table \ref{tab:loss}, for two different pairs of simulations, $p_\text{ms}$, $p_\text{sa}$, and $p_\text{ma}+p_\text{c}$ as calculated by these two approaches typically agree to within 15\%, although $p_\text{ma}$ alone differs by a factor of at least two. This means that the total metal-air interface includes an indeterminate significant fraction of the corners.

We also assumed all surfaces were smooth. Simulations indicate that incorporating smooth bumps on the order of the interface thickness increase the participation ratios and thus loss by a factor of order unity. The value of this factor depends on the interface thickness and on the defect density.

\section{Results for Different Geometries}

Numerical results for a variety of coplanar and microstrip resonator geometries are presented in Table \ref{tab:loss}. We have calculated the participation ratio $p_i$ and loss $p_i\tan\delta$ for a dielectric with thickness 3\,nm, dielectric constant $\epsilon = 10$, and loss tangent $\tan\delta=0.002$, typical values for metal or silicon oxides \cite{Wang2009}.  Since the participation ratio is proportional to thickness, these values can easily be scaled for other parameters. Assuming these parameters, the loss from the metal-air interfaces is typically below $10^{-6}$. The second quantity in the sum for the metal-air columns arises from the 3\,nm by 3\,nm corner at the metal/substrate/air interface. As this is a small area, it shows the sensitivity of the loss to this inside corner and indicates the uncertainty in the metal-air prediction.

\begin{figure}
\begin{center}
\includegraphics[width=3.25in]{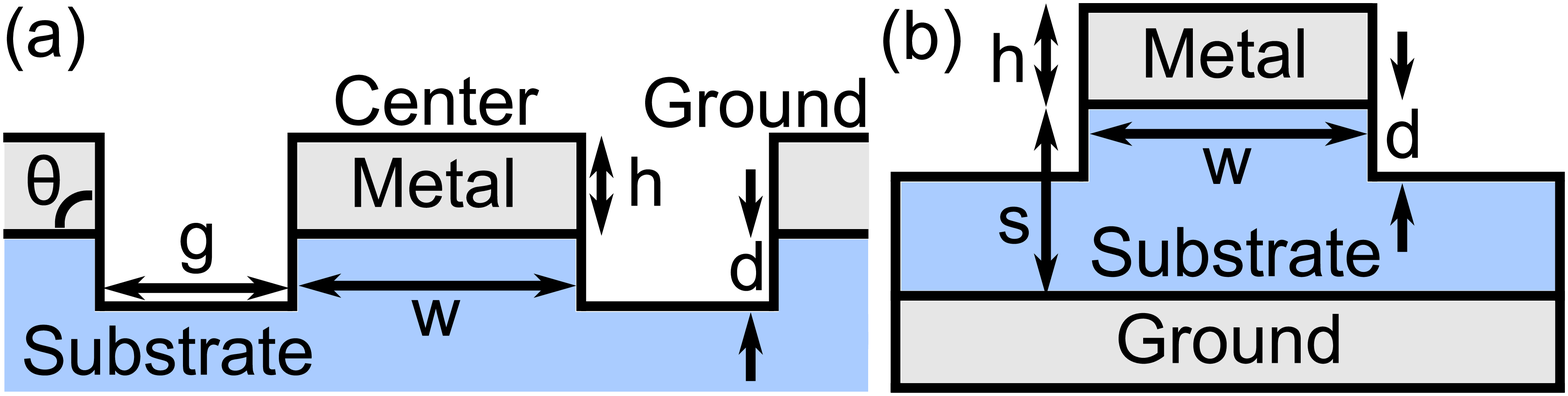}
\end{center}
\caption{Coplanar and microstrip dimensions. (\textbf{a}) Dimensions for coplanar waveguide simulations. We assume $\theta=90^\circ$ unless otherwise specified. (\textbf{b}) Dimensions for microstrip simulations.}
\label{FigDims}
\end{figure}

For coplanar resonators, a significant effect on total loss comes from etching into the substrate within the coplanar gap. This is important because the etching reduces the divergence of the fields at the corner, as shown in Fig.\,2 of the main paper, and because the etching reduces the fields parallel to the metal-substrate interface due to the increased distance from the metal traces. Between the pair of cases c1 and c2 and the pair c3 and c4, it is apparent that $p_\text{ma}$ is reduced by a factor of 2-3. Further deep etching (2\,$\mu$m, case c6) reduces $p_\text{ma}$ by an additional factor of 5 and $p_\text{sa}$ by a factor of 2 while leaving unchanged $p_\text{ms}$. This change has been experimentally tested for Si substrates \cite{Barends2010}, where a feature in the loss versus power saturation curve was identified as substrate-air loss. After etching the substrate caused the feature to disappear and resulted in half the loss, which is consistent with the halving of $p_\text{sa}$ in case c6. This result is consistent with the discussion in the main paper indicating that the substrate-air interface is a dominant loss mechanism.

As experimentally seen by the Delft group \cite{Barends2010}, different geometries of coplanar resonators result in somewhat different losses. In comparing cases c1 and c2, there is minimal difference between the area and surface models except for some change in $p_\text{ma}$ from the inside corner, and comparing cases c4 and c5 indicate that these data are similar to previous work from our group \cite{Wang2009}. The metal thickness is shown to have little effect in case c7. Case c8 shows that decreasing the width makes the loss increase. However, in case c9, increasing the gap from 2 to 20\,$\mu$m gave roughly a factor of 3 reduction in all losses.

Sloped sidewalls are also seen to give different losses by comparing cases c11, c4, and c12, where the metal angle $\theta$ at the substrate-corner corner (Fig.\,\ref{FigDims}) was varied. All interfaces except the metal-air interface had the greatest loss in case c11, where the sidewall slope was $\theta=45^\circ$, and the least loss in case c12, where the overetched sidewalls had $\theta=135^\circ$. This is expected since the electric field is predicted to scale with the distance $r$ from corners as $r^{-3/7}$ for $\theta=45^\circ$, $r^{-1/3}$ for $\theta=90^\circ$, and $r^{-1/5}$ for $\theta=135^\circ$ \cite{Jackson}, thus giving the least field divergence, and thus the lowest loss, at the overetched corner. The metal-air interface exhibits the opposite trend, which is consistent with the same argument for the top corner.

Microstrip resonators show significantly higher capacitance per length, which for the same interface energy results in lower loss. In the base case m1, $p_\text{sa}$ and $p_\text{ma}+p_\text{c}$ are both much less than the corresponding values for coplanar resonators. However, microstrip resonators also have a larger $p_\text{ms}$ contribution, approximately equal to the distance ratio $2t/s$ for oxide thickness $t$. In fact, the thin dielectric of case m2 compared to the base case m1 has a capacitance and $p_\text{ms}$ approximately 10 times that in case m1 and similar $p_\text{sa}$ and $p_\text{ma}$. Another geometric parameter that is important is the width $w$, with which $p_\text{ma}$ and $p_\text{sa}$ scale inversely, as shown by case m3. However, as indicated in case m4, changing the metal height has minimal effect on the participation ratios.

As with coplanar resonators, changing the depth of etching of the exposed substrate also has a significant effect on loss. The lack of dielectric etching in case m5 results in an increase in $p_\text{ma}$ from the inside corner of the metal, which is mostly compensated for by a decrease in $p_\text{sa}$. Case m6 shows that a deeper etch significantly reduces both of these terms. If, instead of etching the exposed dielectric, the dielectric extends up the sidewall of the metal (as in case m7), the losses are similar to that of a deep etch. Overall, the improvements with etching the exposed dielectric for microstrip resonators are primarily at the metal-air interface, but some effect (factor of 2) is seen for the substrate-air interface. However, the most important concern for a microstrip geometry is a low loss metal-substrate interface, as this loss was dominant in all cases.

\section{Derivation of Power Dependence of Participation Ratios}

One potential way to determine which interface dominates the loss is by measuring the power dependence of the loss \cite{Barends2010}. When the saturation of surface two-level states (TLSs) at the field $E_s$ is considered, the surface participation ratio $p$ is given by
\begin{equation}
\frac{p}{t\epsilon} = \int_{r_c}^{r_0} \frac{E^2\,dr}{\sqrt{1+E^2/E_s^2}},
\label{SuppEqSurfaceDef}
\end{equation}
where the interface has thickness $t$ and dielectric constant $\epsilon$. The electric field is assumed to be dominated by a feature such as a corner with length coordinate $r$ from this feature, characteristic length $r_0$, and cutoff length $r_c<r_0$.

For a square corner, the field scales \cite{Jackson} as $E=E_0(r/r_0)^{-1/3}$, so substituting this into Eq.\,(\ref{SuppEqSurfaceDef}) gives a surface participation ratio of
\begin{eqnarray}
\frac{p}{t\epsilon}
&=& E_0^2 \int_{r_c}^{r_0} \frac{(r_0/r)^{2/3}\,dr}{\sqrt{1+(E_0^2/E_s^2)(r_0/r)^{2/3}}} \nonumber \\
&=& 3E_0^2r_0 \left[\sqrt{1+\frac{E_0^2}{E_s^2}} - \sqrt{\left(\frac{r_c}{r_0}\right)^{2/3}+\frac{E_0^2}{E_s^2}}\right].
\end{eqnarray}

For a thin edge at distances much greater than the film thickness, the field scales \cite{Jackson} as $E=E_0(r/r_0)^{-1/2}$, so substituting this into Eq.\,(\ref{SuppEqSurfaceDef}) gives a surface participation ratio of
\begin{eqnarray}
\frac{p}{t\epsilon}
&=& E_0^2 \int_{r_c}^{r_0} \frac{(r_0/r)\,dr}{\sqrt{1+(E_0^2/E_s^2)(r_0/r)}} \nonumber \\
&=& 2E_0^2r_0 \log \frac {1+\sqrt{1+\frac{E_0^2}{E_s^2}}} {\sqrt{\frac{r_c}{r_0}}+\sqrt{\frac{r_c}{r_0}+\frac{E_0^2}{E_s^2}}}.
\end{eqnarray}